\documentclass[journal]{IEEEtran}
\IEEEoverridecommandlockouts
\usepackage{amsmath,amsfonts}
\usepackage{algorithmic}
\usepackage{array}
\usepackage[caption=false,font=footnotesize,labelfont=rm,textfont=rm]{subfig}
\usepackage{textcomp}
\usepackage{stfloats}
\usepackage{url}
\usepackage{verbatim}
\usepackage{graphicx}
\usepackage{balance}
\usepackage{xcolor}
\usepackage{epstopdf}
\usepackage{multirow}
\usepackage{booktabs}
\usepackage{latexsym}
\usepackage{cite}
\usepackage{algorithm}
\usepackage{hyperref}
\usepackage{marvosym}
\usepackage[numbers, sort&compress]{natbib}
\usepackage{nomencl}
\usepackage{textcomp,mathcomp}
\usepackage{color}
\usepackage{enumitem}
\usepackage{mathrsfs}
\usepackage{orcidlink}
\usepackage{amsthm}

\makeatletter
\renewcommand{\@maketitle}{%
\newpage
\centering
{\LARGE  \@title\par} 
\vskip 1.0em
{\large \@author\par} 
\par}
\makeatother

\begin{document}
	
	\title{Decentralized Analysis Approach for Oscillation Damping in Grid-Forming and Grid-Following Heterogeneous Power Systems}

	\author{Xiang Zhu,
    Xiuqiang He,~\IEEEmembership{Member,~IEEE,}
    Hongyang Qing,~\IEEEmembership{Member,~IEEE,}
	and Hua Geng,~\IEEEmembership{Fellow,~IEEE}

	
    }
	
	\maketitle
	
	\begin{abstract}  	

    This letter proposes a decentralized local gain condition~(LGC) to guarantee oscillation damping in inverter-based resource~(IBR)-dominated power systems. The LGC constrains the dynamic gain between each IBR and the network at its point of connection. By satisfying the LGC locally, the closed-loop poles are confined to a desired region, thereby yielding system-wide oscillation damping without requiring global information. Notably, the LGC is agnostic to different IBR dynamics, well-suited for systems with heterogeneous IBRs, and flexible to various damping requirements. Moreover, a low-complexity algorithm is proposed to parameterize LGC, providing scalable and damping-constrained parameter tuning guidance for IBRs.

	\end{abstract}
	
	
	\section{Introduction}

    \IEEEPARstart{P}{ower} systems are shifting from synchronous generator-dominated grids to those dominated by inverter-based resources (IBRs), categorized into grid-forming (GFM) and grid-following (GFL). IBR-dominated systems generally feature weak damping. For homogeneous systems with GFM IBRs only, existing studies typically employed centralized analysis, such as solving the quadratic eigenvalue problem~\cite{Feng},~\cite{Gao}, to constrain the pole distribution for oscillation damping. However, such centralized analysis frameworks suffer from poor scalability to different IBRs, which renders them ineffective for oscillation damping in GFM-GFL heterogeneous systems.

    Decentralized frameworks offer potential for addressing the challenge of GFM-GFL heterogeneity in IBR-dominated systems. Existing works have proposed decentralized analysis methods based on passivity~\cite{Pass} and gain-phase characteristics~\cite{Huang}, which, however, only guarantee that the closed-loop poles lie in the left-half plane, without explicitly satisfying damping ratio requirements. Therefore, guaranteeing oscillation damping in GFM-GFL heterogeneous systems in a decentralized manner remains an unresolved challenge.

    To address this gap, we propose decentralized conditions of whole-system pole distribution for oscillation damping. We reveal the relationship between the \emph{local gain} and the closed-loop pole distribution. Here, the local gain is defined as the gain of dynamic interactions between each IBR and the network at its point of connection, with no restrictions on the dynamics of IBRs. On this basis, decentralized conditions are derived: if all local gains are sufficiently large within a predefined \emph{prohibited domain}, no poles lie within this domain, thus constraining the pole distribution to meet desired damping requirements. Notably, by customizing the prohibited domain, the proposed conditions can flexibly meet various damping requirements. The major contributions of this work are:
	\begin{enumerate}
        \item A novel local gain condition~(LGC) and its low-complexity simplification, the local gain boundary condition (LGBC), are proposed. These conditions ensure fully decentralized and IBR-dynamic-agnostic oscillation damping with flexible damping adaptability.
        
        \item To further parametrize the proposed gain conditions, an efficient algorithm is developed to enable parallel computation of control parameter feasible regions for IBRs. The algorithm provides scalable and damping-constrained parameter tuning guidance for heterogeneous IBRs.
	\end{enumerate}

	\begin{figure}
		\centering
		\includegraphics[width=0.95\linewidth]{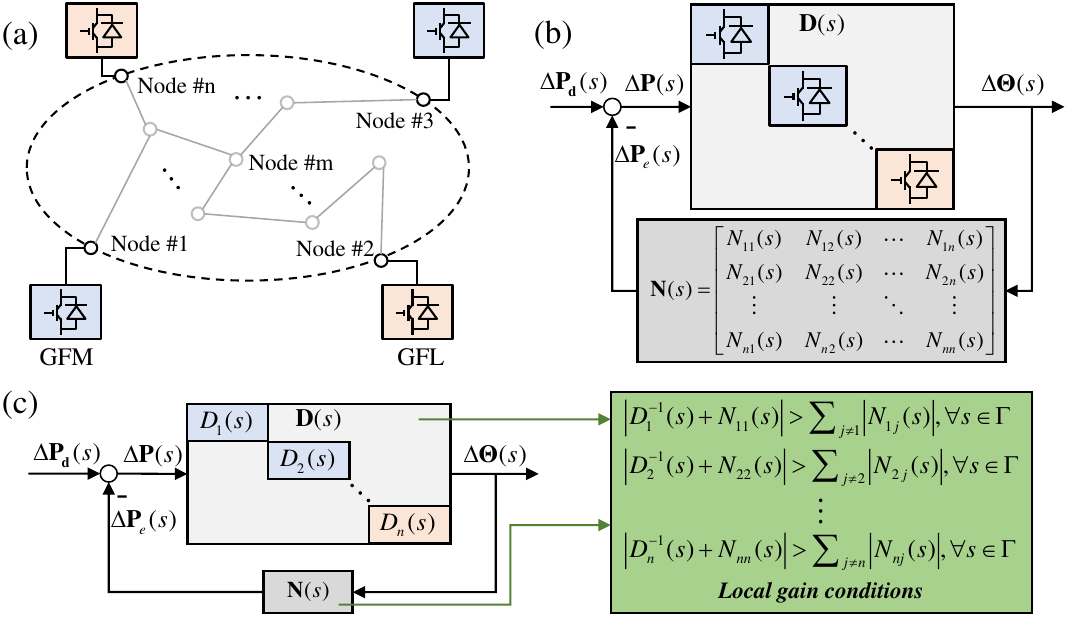}
        \vspace{-0.2cm}
		\caption{(a)~Heterogeneous IBRs are interconnected to form a power system. (b)~Closed-loop system model with feedback connection. (c)~Oscillation damping is achieved by verifying conditions of local gain, i.e., $|D_i(s)+N_{ii}(s)|$, established by each IBR $D_i(s)$ and local network dynamics $N_{ii}(s)$.} 
        \vspace{-0.5cm}
		\label{Sys.Mod} 
	\end{figure}
    
    \section{System Modeling}
    
    We establish a phase-angle-dominated system model for small-signal synchronization stability~\cite{Feng}, which can be extended to include voltage dynamics. The active power-phase angle dynamics of GFL ($D_i^f(s)$) and GFM ($D_i^g(s)$) follow the model in~(\ref{d.1})~\cite{DVPP}, with a compact matrix representation given in~(\ref{d.3}).
    \begin{equation}
        D_i^f(s) = -\frac{{\Delta P_i^f(s)}}{{\Delta \theta _i^f(s)}}, D_i^g(s) = -\frac{{\Delta \theta _i^g(s)}}{{\Delta P_i^g(s)}}
        \label{d.1}
    \end{equation}
    \begin{equation}
        \textbf{D}(s)\left[ {\begin{array}{*{20}{c}}
        {{\bf{\Delta }}{{\textbf{P}}^{\textbf{g}}}}(s)\\
        {{\bf{\Delta }}{{\textbf{P}}^{\textbf{f}}}}(s)
        \end{array}} \right] = {\left[ {\begin{array}{*{20}{c}}
        {{\bf{\Delta }}{{\bf{\Theta }}^{\textbf{g}}}}(s)\\
        {{\bf{\Delta }}{{\bf{\Theta }}^{\textbf{f}}}}(s)
        \end{array}} \right]}
        \label{d.3}
    \end{equation}
    where the vectors ${{\bf{\Delta }}{{\bf{\Theta }}^{\textbf{g}}}}(s)$ and ${{\bf{\Delta }}{{\bf{\Theta }}^{\textbf{f}}}}(s)$ denote the virtual angle dynamics of GFM and GFL and ${{\bf{\Delta}}{{\textbf{P}}^{\textbf{g}}}}(s)$ and ${{\bf{\Delta}}{{\textbf{P}}^{\textbf{f}}}}(s)$ represent their corresponding active power dynamics. 
    
    The relationship between node $i$ and node $j$ is formulated in (\ref{N.1}), where $\rho$ denotes the resistance-inductance ratio, following the topological modeling assumptions provided in \cite{Net}.
    \begin{equation}
        {\Delta  P_{ij}}(s) = \left( {\frac{{{\omega _0}}}{{{s^2} + 2\rho s + \omega _0^2 + {\rho ^2}}}\frac{1}{{{l_{ij}}}}} \right){\Delta \theta _{ij}}(s)
        \label{N.1}
    \end{equation}

    Based on~(\ref{N.1}), a compact matrix representation is derived as $\textbf{Y}(s)\Delta {\bf{\Theta}}(s) = \Delta{\textbf{P}}(s)$. By performing Kron-Reduction on $\textbf{Y}(s)$, the network model is obtained in (\ref{N.2}).
    \begin{equation}
        \textbf{N}(s){\left[ {\begin{array}{*{20}{c}}
        {{\bf{\Delta }}{{\bf{\Theta }}^{\textbf{g}}}}(s)\\
        {{\bf{\Delta }}{{\bf{\Theta }}^{\textbf{f}}}}(s)
        \end{array}} \right]} = \left[ {\begin{array}{*{20}{c}}
        {{\bf{\Delta }}{{\textbf{P}}^{\textbf{g}}_{e}}}(s)\\
        {{\bf{\Delta }}{{\textbf{P}}^{\textbf{f}}_{e}}}(s)
        \end{array}} \right]
        \label{N.2}
    \end{equation}
    where the vectors ${{\bf{\Delta}}{{\textbf{P}}^{\textbf{g}}_{e}}}(s)$ and ${{\bf{\Delta}}{{\textbf{P}}^{\textbf{f}}_{e}}}(s)$ are the active power feedback dynamics of GFM and GFL. Therefore, the system including heterogeneous IBRs~(Fig.~\ref{Sys.Mod}.~(a)) is modeled through feedback interconnection of $\textbf{D}(s)$ and $\textbf{N}(s)$ in Fig.~\ref{Sys.Mod}.~(b).
    
	\section{Gain-Based Decentralized Conditions} \label{sec2}
    The closed-loop characteristic equation of the system is given by~(\ref{T1.1}). The distribution of roots of~(\ref{T1.1}), i.e., the closed-loop poles, corresponds to the damping performance.
    \begin{equation}
         \det \left(\textbf{I}+ \textbf{D}(s)\textbf{N}(s) \right) = 0
         \label{T1.1}
    \end{equation}

    \noindent \textbf{Proposition 1}(Local Gain Condition~(LGC)). \emph{For equation~(\ref{T1.1}), assume that the matrix $\textbf{D}(s)$ to be non-singular within a prohibited domain $\Gamma$. Thus, if the LGC condition~(\ref{T1.2}) is satisfied in $\Gamma$ for each IBR, no closed-loop poles lie within this domain.}
    \begin{equation}
        \left| {D_i^{ - 1}(s) + N_{ii}(s)} \right| > \sum\nolimits_{j \ne i} {\left| {{N_{ij}}}(s) \right|} ,\forall s \in \Gamma
        \label{T1.2}
    \end{equation}
    \begin{proof}
    \vspace{-0.1cm}
        Under the assumption that $\det(\textbf{D}(s)) \ne 0$, the equation (\ref{T1.1}) can be rewritten as (\ref{T1.3}) equivalently. If LGC (\ref{T1.2}) is satisfied for each $D^{-1}_{i}(s)+N_{ii}(s)$, $(\textbf{D}^{-1}(s)+\textbf{N}(s))$ will meet the strict diagonal dominance condition in $\Gamma$.
        \begin{equation}
            \det \left( {{\textbf{D}(s)}} \right)\det \left( \textbf{D}^{-1}(s)+\textbf{N}(s) \right) = 0
            \label{T1.3}
        \end{equation}
        \begin{equation}
            \det \left(\textbf{I}+ \textbf{D}(s)\textbf{N}(s) \right) \ne 0,\forall s \in \Gamma 
            \label{T1.6}
        \end{equation}
        
        According to \emph{Levy-Desplanques Theorem}~\cite{Levy}, if $(\textbf{D}^{-1}(s)+\textbf{N}(s))$ satisfies the strict diagonal dominance condition in $\Gamma$, then it's non-singular in $\Gamma$, thus~(\ref{T1.6}) is satisfied.   
        \vspace{-0.2cm}
    \end{proof}
    \noindent \textbf{Remark}. \emph{Physically, the LGC mandates that the dynamic interactions at each point of connection must strictly exceed the coupling dynamics from the external networks. In this way, the local damping capabilities are sufficient to counteract oscillations propagating through network interconnections.}
    
    Practically, the direct solution of the LGC-derived complex-function modulus inequality involves a high computational burden, motivating the proposal of a sufficient condition.
    
    \noindent \textbf{Proposition 2}~(Local Gain Boundary Condition~(LGBC)). \emph{Assume that $\textbf{D}^{-1}(s)$ and $\textbf{N}(s)$ are analytical within $\Gamma$. If each diagonal element of $(\textbf{D}^{-1}(s)+\textbf{N}(s))$ is non-zero in $\Gamma$, then verification of the LGC for $(\textbf{D}^{-1}(s)+\textbf{N}(s))$ only needs to be performed on the boundary of $\Gamma$, i.e., $\mathcal{B}$, as shown in~(\ref{T2.1}):}    
    \begin{equation}
        \left\{ \begin{array}{l}
        D_i^{ - 1}(s) + {N_{ii}}(s) \ne 0,\forall s \in \Gamma \\
        {\left| {D_i^{ - 1}(s) + {N_{ii}}(s)} \right|} > \sum\nolimits_{j \ne i}{\left| {{N_{ij}}}(s) \right|} ,\forall s \in \mathcal{B}
        \end{array} \right.
        \label{T2.1}
    \end{equation}
    \begin{proof}
    \vspace{-0.2cm}
        Under the aforementioned assumption, if $D^{-1}_{i}(s) + N_{ii}(s) \neq 0$ for all $s \in \Gamma$, then $\frac{1}{D^{-1}_{i}(s) + N_{ii}(s)}$ is analytic within the domain $\Gamma$. By virtue of the \emph{Maximum Modulus Principle}~\cite{MMP}, the maximum of $\left| \frac{1}{D^{-1}_{i}(s) + N_{ii}(s)} \right|$ (corresponding to the minimum of $\left| D^{-1}_{i}(s) + N_{ii}(s) \right|$) and the maximum of $\sum_{j \ne i} \left| N_{ij}(s) \right|$ must be attained on the boundary of $\Gamma$. Consequently, if the inequality $\min\left|D^{-1}_{i}(s)+N_{ii}(s)\right| > \max \sum\nolimits_{j \ne i}{\left| {{N_{ij}}}(s) \right|}$ holds on $\mathcal{B}$, it strictly guarantees satisfaction of LGC within $\Gamma$.
    \end{proof}
    \vspace{-0.2cm}

    Furthermore, given that dynamics of IBRs and networks are usually strictly proper rational functions, as $|s| \to \infty$, $\left| D^{-1}_{i}(s) + N_{ii}(s) \right| \to \infty$ and $\left| N_{ij}(s) \right| \to 0$ hold uniformly. This implies that the minimum of $\left| D^{-1}_{i}(s) + N_{ii}(s) \right|$ and the maximum of $\sum_{j \ne i} \left| N_{ij}(s) \right|$ can only be attained on the finite boundary of $\Gamma$, i.e., $\mathcal{B}^f$. Consequently, the LGBC significantly reduces the computational burden of verifying the LGC. Both the LGC and LGBC are generalized conditions that impose no restrictions on the specific dynamics of IBRs, and are therefore inherently IBR-dynamic-agnostic.

	\section{Parameter Feasible Regions Determination} \label{sec3}
    
    \subsection{Definition of the Prohibited Domain}
    
	\begin{figure}
    \vspace{-0.3cm}
		\centering
		\includegraphics[width=0.9\linewidth]{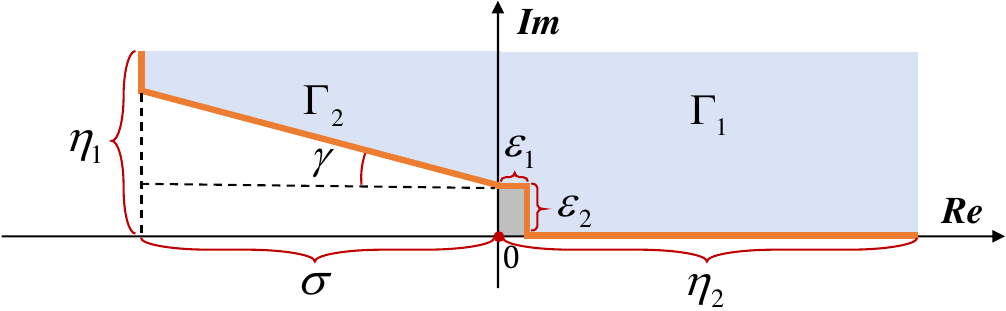}
		\caption{The diagram of the prohibited domain and its boundary. The vertical boundary, $\text{Re}(s)=-\sigma$, facilitates the application of the Routh-Hurwitz criterion to derive the satisfaction condition for $D_i^{ - 1}(s) + {N_{ii}(s)} \ne 0$. Specifically, when the network matrix $\textbf{N}(s)$ takes the form of a Laplacian matrix, the system inherently exhibits a pole at the origin~\cite{Net}.} 
        \vspace{-0.5cm}
		\label{Forbid} 
	\end{figure} 
    Without loss of generality, the prohibited domain $\Gamma$ is defined in~(\ref{FP.1}) based on the symmetry of closed-loop pole distribution. It consists of two regions: $\Gamma_1$, the closed right-half plane excluding the origin, which ensures system asymptotic stability; and $\Gamma_2$, the left-half plane region adjacent to the imaginary axis and characterized by a predefined damping ratio ($\xi = \cos \gamma$), which prevents weakly damped oscillations. Additionally, the design of $\Gamma$ is unrestricted and can be flexibly customized to accommodate diverse engineering requirements for oscillation damping.
    \begin{equation}
        \begin{array}{l}
        \Gamma  = {\Gamma _1} \cup {\Gamma _2}\\
        \left\{ \begin{array}{l}
        {\Gamma _1} = \left\{ {\left. s \right|{\mathop{\rm Re}\nolimits} \left( s \right) \ge 0,{\mathop{\rm Im}\nolimits} \left( s \right) \ge 0,s \ne 0} \right\}\\
        {\Gamma _2} = \left\{ {\left. s \right| - \sigma  \le {\mathop{\rm Re}\nolimits} \left( s \right) < 0, {\frac{{{\mathop{\rm Im}\nolimits} \left( s \right)}}{{ - {\mathop{\rm Re}\nolimits} \left( s \right)}} \ge \tan \gamma } } \right\}
        \end{array} \right.
        \end{array}
        \label{FP.1}
    \end{equation}
    
    As illustrated in Fig.~\ref{Forbid}, and considering the origin, the finite boundary $\mathcal{B}^f$ corresponding to the prohibited domain $\Gamma$ can be expressed as given in~(\ref{FP.2}), where the smaller the values of $\varepsilon_1$ and $\varepsilon_2$, the higher the accuracy of the boundary $\mathcal{B}^f$.
    \begin{equation}
		  \left\{ \begin{aligned}
			& {\mathop{\rm Re}\nolimits} \left( s \right) =  - \sigma ,\sigma \tan \gamma+\varepsilon_2  \le {\mathop{\rm Im}\nolimits} \left( s \right) \le {\eta _1} \\
			& - \sigma  \le {\mathop{\rm Re}\nolimits} \left( s \right) <  0 ,{\mathop{\rm Im}\nolimits} \left( s \right) = \varepsilon_2 - {\mathop{\rm Re}\nolimits} \left( s \right)\tan \gamma\\
			& 0 \le {\mathop{\rm Re}\nolimits} \left( s \right) \le \varepsilon_1 ,{\mathop{\rm Im}\nolimits} \left( s \right) = \varepsilon_2\\
			& {\mathop{\rm Re}\nolimits} \left( s \right) = \varepsilon_1, 0 \le {\mathop{\rm Im}\nolimits} \left( s \right) \le \varepsilon_2 \\
            & \varepsilon_1  \le {\mathop{\rm Re}\nolimits} \left( s \right) \le {\eta _2} ,{\mathop{\rm Im}\nolimits} \left( s \right) = 0
		\end{aligned} \right.
		\label{FP.2}
	\end{equation}
    \subsection{Algorithm for Parameter Feasible Regions Determination}
    Based on the LGBC in~(\ref{T2.1}), we propose an algorithm for determining the control parameter feasible regions oriented towards heterogeneous IBRs, encompassing both GFM and GFL (Algorithm~\ref{alg.1}). The inputs are the finite boundary $\mathcal{B}^f$ of the prohibited domain defined in~(\ref{FP.1}), the device matrix $\textbf{D}(s)$, and the network $\textbf{N}(s)$. Given the decentralized advantage of the LGBC, Algorithm~\ref{alg.1} enables parallel computation of the feasible region $\mathcal{F}_i$ for each IBR, with its computational burden remaining relatively stable regardless of the IBR scale.
    \begin{figure}[htbp]
    \vspace{-0.2cm}
		\renewcommand{\algorithmicrequire}{\textbf{Input:}}
		\renewcommand{\algorithmicensure}{\textbf{Output:}}
		\begin{algorithm}[H]
			\caption{Parameter feasible region determination.}
			\begin{algorithmic}[1]
				\REQUIRE {Boundary $\mathcal{B}^f$, device matrix $\textbf{D}(s)$ and network $\textbf{N}(s)$.}
				\ENSURE {Parameter feasible regions $\mathcal{F} = \{\mathcal{F}_1, \dots, \mathcal{F}_n\}$.} 
                \STATE{Calculate parameter ranges $\{\Xi_i\}_{i=1}^n$ that ensure $\left| D_i^{-1}(s,\Xi_{i}) + N_{ii}(s) \right|\ne 0, \text{Re}(s)>-\sigma$.}
                \STATE {Discretize boundary $\mathcal{B}^f$ into point set $\mathcal{S}$.}
                \FOR{device $i \leftarrow 1$ to $n$}
                    \STATE {$\mathcal{F}_i \leftarrow \emptyset$ and discretize $\Xi_i$ into candidate set $\Psi_i$.}
                    \FOR{each parameter $\psi \in \Psi_i$.}
                        \IF{$\left| D_i^{-1}(s, \psi) + N_{ii}(s) \right| > \sum\nolimits_{j \ne i}|N_{ij}(s)|, \forall s\in\mathcal{S}$.}
                        \STATE{$\mathcal{F}_i \leftarrow \mathcal{F}_i \cup \{\psi\}$.}
                        \ENDIF
                    \ENDFOR
                \ENDFOR  
                \RETURN{$\mathcal{F}$}
			\end{algorithmic}
			\label{alg.1}
		\end{algorithm}
        \vspace{-0.7cm}
	\end{figure} 
	\section{Case Studies} \label{sec4}
    In case studies, we primarily focus on low-frequency oscillation, which is a common oscillation type in power systems~\cite{Feng},~\cite{Gao}. Accordingly, the network model (\ref{N.1}) can be simplified to the linearized static power flow equations, and the dynamics of GFM and GFL are defined as in (\ref{d.4}), where $m_i$ and $H_i$ denote virtual inertia parameters, and $D_i$ and $d_i$ represent virtual damping parameters. $K_i^P$ and $K_i^I$ are the proportional and integral gains of the phase-locked loops. While the network and device models adopt standard and simplified formulations, the proposed conditions are general enough to apply to other models. Algorithm~\ref{alg.1} is utilized to compute the feasible regions of GFM parameters $m_i$ and $d_i$, as well as GFL parameters $H_i$ and $D_i$.
    \begin{subequations}
    \begin{align}
        & D_i^g(s) =  - \frac{1}{{s\left( {{m_i}s + {d_i}} \right)}} \\
        & {\left( {D_i^f(s)} \right)^{ - 1}} =  - \frac{{{s^2} + V_0K_i^Ps + V_0K_i^I}}{{s\left( {{D_i} + {H_i}s} \right) \left( {V_0K_i^Ps + V_0K_i^I} \right)}}
    \end{align}
    \label{d.4}
    \end{subequations}
    
\begin{figure}
		\centering
		\includegraphics[width=0.89\linewidth]{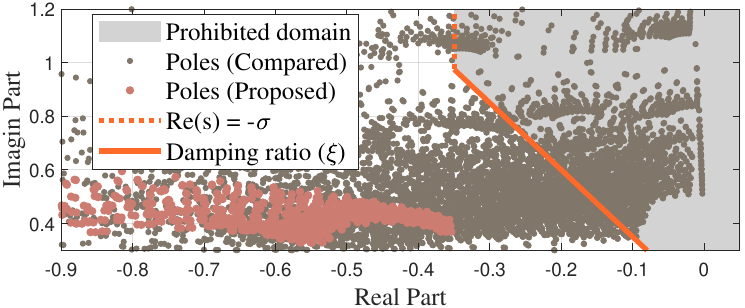}
		\vspace{-0.2cm}
        \caption{Distribution of closed-loop poles~($\text{Im}(s)\ge 0$) of the two-IBR system.}
		\label{Sim.poles}
        \vspace{-0.2cm}
	\end{figure}

    The ranges of $m_i$ and $H_i$ are 0.1 s to 20 s, while those of $d_i$ and $D_i$ are 0.1 pu to 20 pu. The remaining parameters are chosen as: $K^P_1 = 4$, $K^I_1 = 40$, $K^P_2 = 2$, $K^I_2 = 20$, $\sigma = 0.35$, $\xi=\cos \gamma = 0.37$, $\varepsilon_1 = 10^{-3}$, $\varepsilon_2 = 10^{-1}$, and $\eta_1 = \eta_2 = 10$. The voltage setpoints $V_0$ are uniformly set to 1.0 pu. Solving the quadratic equations $s^2 + K_i^Ps + K_i^I = 0$ confirms that $\det(\textbf{D}(s)) \neq 0$ and $D^{-1}(s)$ is analytical within $\Gamma$. A two-IBR test system (one GFM IBR, one GFL IBR) is used to verify the proposed conditions, with pole distributions of Algorithm~\ref{alg.1}-derived~(Proposed) and random configurations~(Compared) shown in Fig.~\ref{Sim.poles}. It is shown that the Algorithm~\ref{alg.1} ensures no poles lie within the predefined prohibited domain, suppressing weakly damped oscillations.

\begin{figure}
		\centering
        \vspace{-0.2cm}
		\includegraphics[width=0.9\linewidth]{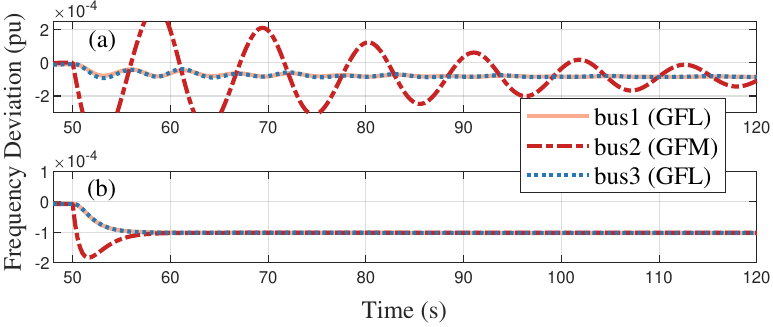}
        \vspace{-0.2cm}
		\caption{Time-domain simulations on the IEEE 9-bus system~\cite{DVPP}: one GFM IBR at bus 2, two GFL IBRs at bus 1 and bus 3, with step power disturbances applied at bus 2 at 50 s. (a) Simulations with non-algorithm parameters. (b) Simulations with Algorithm~\ref{alg.1}-derived parameters.} 
		\label{Sim.results}
        \vspace{-0.6cm}
	\end{figure}

    For a three-IBR system (one GFM IBR, two GFL IBRs), theoretical analysis shows that a parameter configuration failing to satisfy Algorithm~\ref{alg.1}-derived feasible regions yields dominant poles at $-0.078\pm0.627j$ (damping ratio $\xi = 0.12$), which lie within the prohibited domain. Implemented in a MATLAB/Simulink-simulated IEEE 9-bus system, this configuration induces weakly damped oscillations (see Fig.~\ref{Sim.results}(a)). In contrast, a specific parameter configuration obtained via Algorithm~\ref{alg.1} achieves a much higher damping ratio ($\xi = 0.98$), and simulations confirm effective oscillation damping under the same disturbance (see Fig.~\ref{Sim.results}(b)). 
    
    For scalability testing, the feasible regions of each IBR are solved in parallel across different test systems. For the IEEE 9-bus (3 IBRs), 39-bus (10 IBRs), and 118-bus (54 IBRs) systems, the respective computational times for solving parameter feasible regions are 0.571 s, 0.449 s, and 0.526 s. These times are closely comparable, which confirms that the Algorithm~\ref{alg.1} is highly efficient and scalable for calculating control parameter feasible regions of IBRs.

	\section{Conclusion} \label{sec5}

    This letter proposes decentralized conditions to guarantee oscillation damping in systems with heterogeneous IBRs, meeting diverse damping requirements flexibly while remaining agnostic to IBR dynamics. An efficient algorithm is developed to determine the feasible regions for control parameters of IBRs, serving as a practical tool for damping-constrained parameter tuning. Future work will incorporate voltage dynamics to establish a unified framework of decentralized damping conditions, thereby informing the formulation of device-side grid codes tailored for IBR-dominated power systems.
    
	\footnotesize
	\bibliography{reference}
	\bibliographystyle{IEEEtran}

\end{document}